\documentclass[%
 reprint,
superscriptaddress,
 amsmath,amssymb,
 aps,
pra,
]{revtex4-2}

\usepackage{multirow}
\usepackage[T1]{fontenc}
\usepackage[colorlinks, linkcolor=red, anchorcolor=red, citecolor=red]{hyperref}
\usepackage{amsthm}
\usepackage{graphicx}
\usepackage{dcolumn}
\usepackage{bm}
\usepackage{float}
\usepackage{algpseudocodex}
\usepackage[linesnumbered, boxed, ruled]{algorithm2e}
\usepackage{caption}
\usepackage{subcaption}
\usepackage{xspace}
\usepackage{braket}
\usepackage{booktabs}
\usepackage{threeparttable}
\usepackage{quantikz}
\usepackage{tikz,amsmath, amssymb,bm,color}
\usetikzlibrary{calc}
\usepackage{makecell}

\newcommand{\CommentLine}[1]{\Statex \hspace*{\algorithmicindent} \textit{\(\triangleright\) #1}}

\newcommand{\casql}{Laboratory of Quantum Information, University of Science and Technology of China, Hefei 230026, China}

\newcommand{\aihf}{Institute of Artificial Intelligence, Hefei Comprehensive National Science Center, Hefei, Anhui, 230088, China}

\newcommand{\origin}{Origin Quantum Computing, Hefei, Anhui, 230088, China}
\newcommand{\IAT}{Institute of the Advanced Technology, University of Science and Technology of China, Hefei, Anhui, 230088, China}

\newtheorem{theorem}{Theorem}[section]

\newtheorem{lemma}[theorem]{Lemma}

\begin{document}

\preprint{APS/123-QED}


\title{Towards Fault-Tolerant Quantum Deep Learning: Designing and Analyzing Quantum ResNet and Transformer with Quantum Arithmetic and Linear Algebra Primitives}

\author{Xiao-Fan Xu}
\thanks{These authors contributed equally to this work.}
\affiliation{\casql}

\author{Cheng Xue}
\thanks{These authors contributed equally to this work.}
\affiliation{\aihf}

\author{Xi-Ning Zhuang}
\affiliation{\casql}
\affiliation{\origin}

\author{Yun-Jie Wang}
\affiliation{\IAT}

\author{Tai-Ping Sun}
\affiliation{\casql}


\author{Yu Fang}
\affiliation{\aihf}
\affiliation{\IAT}

\author{Jun-Chao Wang}
\affiliation{Laboratory for Advanced Computing and Intelligence Engineering, Zhengzhou 450001, China}

\author{Huan-Yu Liu}
\affiliation{\IAT}

\author{Yu-Chun Wu}
\thanks{wuyuchun@ustc.edu.cn}
\affiliation{\casql}
\affiliation{\IAT}
\author{Zhao-Yun Chen}
\thanks{chenzhaoyun@iai.ustc.edu.cn}
\affiliation{\aihf}

\author{Guo-Ping Guo}
\affiliation{\casql}
\affiliation{\aihf}
\affiliation{\origin}

\begin{abstract}
Achieving a practical quantum speedup for deep neural networks (DNNs) remains a central yet elusive goal, hindered by the dual challenges of constructing deep architectures and the prohibitive overhead of data loading and measurement. We introduce a framework to overcome these barriers, specifically targeting an asymptotic speedup with respect to the large input dimensions of modern DNNs (e.g., sequence length or image size). Our framework enables the design of multi-layer Quantum ResNet and Quantum Transformer models by strategically decomposing tasks: computationally intensive operations on the large input dimension are assigned to quantum linear algebra subroutines, while operations on the smaller, fixed feature dimension are handled by efficient quantum arithmetic. A cornerstone of our approach is a novel data transfer protocol, Discrete Chebyshev Decomposition (DCD), which facilitates this modularity. Numerical validation reveals a pivotal insight: the measurement cost required to maintain a target accuracy scales sublinearly with the input dimension. This sublinear scaling is the key to preserving the quantum advantage, ensuring that I/O overhead does not nullify the computational gains. A rigorous resource analysis further corroborates the superiority of our models in both efficiency and flexibility. Powered by this targeted acceleration strategy and the efficiency of DCD, our framework establishes a viable path toward scalable quantum deep learning.
\end{abstract}

\maketitle
\section{Introduction}

\begin{figure*}[ht]
    \centering
    \includegraphics[width=0.9\linewidth]{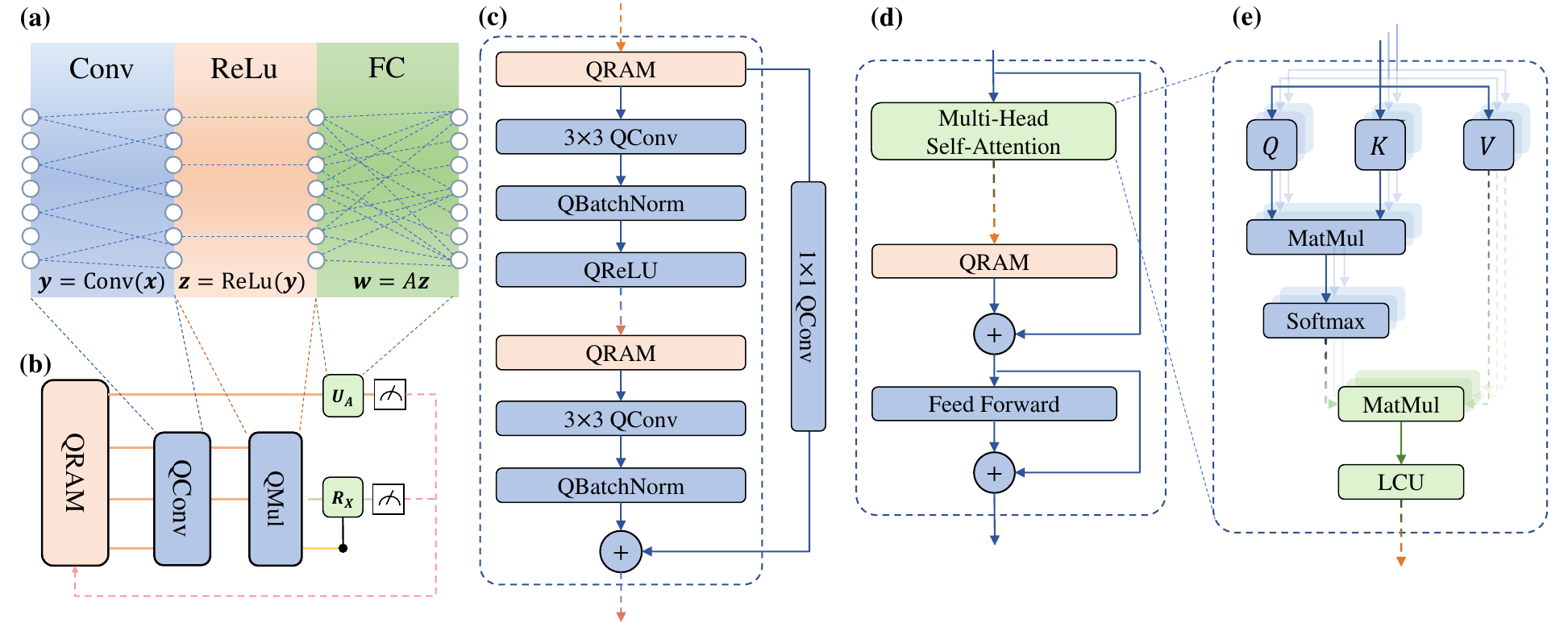}
    \caption{An overview of our hybrid quantum-classical framework for building deep quantum neural networks. 
    The figure first illustrates the direct mapping from \textbf{(a)} a classical tensor operation to \textbf{(b)} its equivalent quantum circuit implementation. In \textbf{(b)}, different lines correspond to the qubits involved in the computation over time; the termination of some lines illustrates the 'uncomputing' process within a Quantum Arithmetic Module (QAM), which allows qubits to be released from the computation. 
    \textbf{(c)}, \textbf{(d)}, and \textbf{(e)} then describe the high-level workflow using a modular block-diagram representation. Our framework is composed of three core module types: Quantum Arithmetic Modules (QAM, blue blocks), Quantum Linear Algebra Modules (QLAM, green blocks), and Data Transfer Modules (DTM, orange blocks). The flow of data associated with a module is represented by arrows of the corresponding color, while the dashed lines indicate the direction of data conversion between different stages.
    These fundamental modules are assembled to construct sophisticated architectures, such as \textbf{(c)} the building block of a Quantum ResNet and \textbf{(d)} the building block of a Quantum Transformer. \textbf{(e)} provides a detailed schematic of the multi-head attention mechanism, demonstrating its concrete implementation using our proposed modules.}
    \label{fig:framework_overview}
\end{figure*}
The current era of artificial intelligence is defined by the triumph of deep neural networks (DNNs). Large-scale models, particularly the Transformer architecture~\cite{vaswani2017attention}, have revolutionized countless fields by leveraging immense depth to learn complex data representations. This success, however, is a double-edged sword. The computational demands of these models are creating a formidable bottleneck, especially for operations whose complexity scales polynomially with the primary \textbf{input dimension}, such as the $O(N^2)$ attention mechanism in Transformers with sequence length $N$. In parallel, quantum computing offers a new paradigm promising significant speedups for such computational tasks~\cite{nielsen2010quantum, preskill2018quantum}. This has catalyzed a surge of research into Quantum Deep Neural Networks (QDNNs)~\cite{beer2020training, liu2024towards, li2020quantum, Kerenidis2020Quantum, ye2025quantum}, aiming to harness quantum mechanics to transcend these specific scaling limitations of classical deep learning.

However, the pursuit of practical QDNNs has splintered into two main directions, each with fundamental limitations. Variational Quantum Circuits (VQCs)~\cite{cerezo2021variational} are compatible with near-term hardware but generally lack provable speedups and are plagued by trainability issues like barren plateaus~\cite{mcclean2018barren,wang2021noise,anschuetz2022quantum,bittel2021training}. Conversely, approaches based on Quantum Linear Algebra (QLA) subroutines~\cite{childs2017quantum,liu2021efficient,krovi2023improved, liao2024gpt} hold the promise of demonstrable polynomial speedups. Yet, these QLA-based methods confront a critical challenge: constructing genuinely deep architectures. The constraints of unitary evolution and the prohibitive cost of inter-layer data transfer have largely confined these proposals to theoretical constructs, unable to offer a resource-efficient path toward deep networks that can effectively tackle large-scale input data.

This work confronts this challenge with a new strategic approach. We posit that a viable quantum advantage can be achieved by selectively accelerating only the parts of a DNN that are bottlenecks with respect to the large input dimension ($N$), while handling operations on the smaller, fixed feature dimension ($d$) with different, more flexible quantum routines. This insight leads us to a novel framework for constructing deep QNNs, built upon a hybrid quantum-classical, layer-by-layer execution model. We employ a modular circuit design that strategically allocates computational tasks involving:
\begin{itemize}
\item\textbf{Quantum Linear Algebra Modules (QLAMs)} are reserved for the dense, large-scale operations that scale with $N$;
\item\textbf{Quantum Arithmetic Modules (QAMs)} are used for element-wise non-linearities and structured operations on the small dimension $d$.
\item\textbf{Data Transfer Modules (DTMs)} facilitate the transfer of data between the classical and quantum domains, encompassing key processes such as quantum measurement, state preparation, and Quantum Random Access Memory (QRAM).
\end{itemize}
The key to enabling this deep, modular architecture is our ``good-enough'' information transfer principle, embodied by a data transfer protocol we term \textbf{Discrete Chebyshev Decomposition (DCD)}. DCD efficiently extracts a compressed classical representation between layers, making the entire hybrid model feasible by overcoming the data I/O bottleneck.

In this work, we provide the first concrete theoretical and empirical validation of this targeted acceleration strategy as a viable pathway toward large-scale QDNNs. Our key contributions are as follows:
\begin{itemize}
    \item \textbf{Targeted Acceleration Framework and Models:} We design and detail the first concrete multi-layer qResNet and qTransformer models built on a strategy of selective quantum acceleration. We establish the theoretical foundation for their end-to-end speedup with respect to the primary input dimension.
    \item \textbf{Validation of Sublinear Scaling:} We demonstrate through numerical experiments a crucial sublinear scaling law: the measurement cost required to maintain a target accuracy scales sublinearly with the input dimension. This provides strong quantitative evidence that the I/O overhead does not negate our targeted computational speedup.
    \item \textbf{Novel Protocol and Intrinsic Structure Discovery:} We introduce the DCD protocol for efficient, ``good-enough'' data transfer. We further discover that DCD reveals an unexpectedly compact, dataset-independent feature representation within the qTransformer, pointing to an intrinsic low-rank structure that can be exploited for further optimization.
    \item \textbf{Rigorous Resource Analysis:} We provide a comprehensive resource analysis, quantitatively demonstrating that our models offer a significant practical advantage over state-of-the-art proposals and marking a critical step from theoretical possibility to practical realization.
\end{itemize}

\section{A Framework for Deep Quantum Networks}
\label{sec:framework}

The primary obstacles to realizing deep quantum networks are the constraints imposed by the no-cloning theorem and the prohibitive resource cost of high-fidelity quantum state transfer between layers. To overcome these challenges, we introduce a practical and modular framework for deep quantum models built upon a foundational principle: a deep network can achieve high performance without perfect, high-fidelity reconstruction of its intermediate quantum states. This ``good-enough'' principle, which posits that preserving salient features is more critical than exact state replication, enables a modular, layer-by-layer architecture. While this design philosophy has been considered in other quantum machine learning contexts~\cite{sharma2022trainability, perez2020data}, our work provides the first concrete, resource-efficient implementation for deep, QLA-based models.

\subsection{The Quantum-Classical Interface Bottleneck}

A central challenge in constructing deep quantum models is managing the flow of information between layers. A purely quantum, multi-layer architecture faces formidable obstacles. The no-cloning theorem forbids copying an unknown quantum state, preventing the straightforward implementation of architectures with fan-out connections. Moreover, non-unitary operations like non-linear activations necessitate measurement and state re-preparation, breaking quantum coherence and leading to circuit complexity that can grow exponentially with network depth.

This reality compels a hybrid, layer-by-layer execution model, where one quantum layer's output is measured before configuring the next. However, this introduces a new, severe bottleneck: the quantum-to-classical (Q2C) data transfer. The standard method for this task, Quantum State Tomography (QST), reconstructs a state's density matrix but requires a number of measurements that scales polynomially with the Hilbert space dimension $d$ (e.g., $O(d^2)$), rendering it impractical. While advanced techniques like classical shadow tomography~\cite{huang2020predicting} reduce this scaling for specific tasks, the overhead for reconstructing a high-fidelity state vector remains a fundamental barrier. This high I/O cost can easily overwhelm any polynomial speedup offered by internal quantum subroutines, crippling the entire model. Overcoming this interface bottleneck is therefore not merely an optimization but a prerequisite for achieving any practical quantum advantage in deep learning.

\subsection{Hybrid Architecture via ``Good-Enough'' State Re-preparation}
\label{sec:hybrid_arch}

The foundation of our framework is the insight that effective performance, not perfect fidelity, is the key to scalability. Classical deep learning has long shown that networks are resilient to perturbations like quantization and pruning, as their efficacy hinges on salient features, not high-precision values~\cite{han2015learning}. We operationalize this ``good-enough'' principle in the quantum domain through a novel inter-layer connection mechanism, realized within a hybrid quantum-classical, layer-by-layer architecture (Figure~\ref{fig:framework_overview}).

Instead of a single, monolithic quantum circuit, we treat each layer as a self-contained computational unit. A forward pass through the network proceeds iteratively:
\begin{enumerate}
    \item \textbf{Classical-to-Quantum (C2Q):} Classical data describing the input for layer $k$ is encoded into an initial quantum state $\ket{\psi_{\text{in}}^{(k)}}$.
    \item \textbf{Quantum Execution:} The quantum circuit for layer $k$, composed of QLAM and QAM modules, is executed to produce the output state $\ket{\psi_{\text{out}}^{(k)}}$.
    \item \textbf{Quantum-to-Classical (Q2C):} An efficient measurement protocol—the core of our contribution—extracts a \textit{compressed classical representation} of $\ket{\psi_{\text{out}}^{(k)}}$.
    \item \textbf{Classical Processing \& Iteration:} This classical vector is used to configure layer $k+1$, and the loop repeats.
\end{enumerate}
This partitioned strategy effectively decouples the quantum resources of adjacent layers, controlling the accumulation of circuit depth and mitigating error propagation. The viability of this entire framework hinges on a Q2C protocol that is efficient enough to preserve the overall quantum advantage.

\subsection{Intra-Layer Computation: Modular Quantum Circuits}
\label{sec:intra_layer}

Our framework's design is tailored for a common and critical regime in deep learning: where the primary input dimension is significantly larger than the internal feature dimension. For instance, in vision transformers, the sequence length $N$ (number of patches) is often much greater than the embedding dimension $d$ ($N \gg d$); in convolutional networks, the image spatial dimensions $H, W$ are much larger than the channel count $C$. Our goal is to achieve a quantum speedup with respect to these large input dimensions. This principle dictates our choice of quantum modules for intra-layer computation.

\subsubsection{Quantum Linear Algebra Module (QLAM)}
\label{sec:qlam}
The QLAM is reserved for operations that are computationally dense with respect to the large input dimension (e.g., $N$). It is a high-level abstraction for block-encoding-based algorithms designed to tackle the primary computational bottlenecks of a model, such as:
\textbf{Large-Scale Matrix Multiplication:} Performing products of matrices whose dimensions scale with $N$, such as the $N \times N$ attention score matrix in a Transformer;
and \textbf{Inner Product Estimation:} Given state preparation oracles for vectors $\ket{\psi_x}$ and $\ket{\psi_y}$, the QLAM efficiently estimates their squared overlap, $|\langle \psi_x | \psi_y \rangle|^2$, via measurement. This is crucial for applications in backpropagation.

\subsubsection{Quantum Arithmetic Module (QAM)}
\label{sec:qam}
The QAM complements the QLAM by handling computations that are either element-wise or can be structured as parallel operations on the smaller, fixed feature dimension (e.g., $d$). It operates on digitally encoded numbers and is essential for:
\textbf{Element-wise Non-linearities:} Applying functions like ReLU or approximate softmax arithmetically;
\textbf{Structured Linear Algebra:} Performing linear operations that act independently on each of the $N$ input items. For example, multiplying an $N \times d$ input matrix by a $d \times d$ weight matrix can be viewed as $N$ parallel $d$-dimensional vector-matrix products, a task well-suited for the QAM's arithmetic capabilities;
and \textbf{Sparse Operations:} Executing structured sparse operations like convolutions.

\subsubsection{Module Composition and Data Flow}
\label{sec:composition}
A typical layer is constructed by composing QLAM and QAM instances. For example, a dense layer with ReLU activation proceeds as follows:
\begin{enumerate}
    \item An input state $\ket{\psi_{\text{in}}}$ is processed by a \textbf{QLAM} (with weight matrix $W$) to compute the matrix-vector product.
    \item The output is passed to a \textbf{DTM} to reconstruct the quantum data.
    \item The reconstructed data is passed to a \textbf{QAM} to arithmetically add a bias vector.
    \item A second \textbf{QAM} applies the ReLU activation function element-wise.
    \item The final state, $\ket{\psi_{\text{out}}}$, is ready for Q2C extraction.
\end{enumerate}
This modularity requires efficient conversion between the amplitude encoding used by QLAM and the digital encoding required by QAM, a key design challenge addressed using routines based on Quantum Random Access Memory (QRAM) and amplitude estimation.

\subsection{Inter-Layer Communication: The Discrete Chebyshev Decomposition Protocol}
\label{sec:dcd}
The critical link in our architecture—and the concrete embodiment of our ``good-enough'' principle—is the protocol for Q2C and C2Q conversion. To bypass the infeasible cost of tomography, we introduce the Discrete Chebyshev Decomposition (DCD) protocol, designed to extract a compressed classical representation of the quantum state.

\subsubsection{Theoretical Foundation}
Our choice of the Chebyshev basis is mathematically motivated. For any function on a finite interval, a truncated Chebyshev series provides the best polynomial approximation in the $l_\infty$ norm (minimax approximation). In our context, the amplitudes of a quantum state $\ket{\psi}$ can be viewed as evaluations of an underlying function. By projecting $\ket{\psi}$ onto the first $r$ Chebyshev basis vectors, we are effectively finding the optimal low-degree polynomial approximation of this function. This captures its most significant, low-frequency features with a minimal number of coefficients, providing maximum information for a given level of compression.

\subsubsection{Algorithm Description}
The DCD protocol assumes that the information in a layer's output state $\ket{\psi}$ is highly compressible. Any such state can be formally expanded in the discrete Chebyshev basis $\{\ket{T_j}\}_{j=0}^{d-1}$ as $\ket{\psi} = \sum_{j=0}^{d-1} c_j \ket{T_j}$, where $c_j = \braket{T_j | \psi}$. Our core hypothesis is that an approximation using only the first $r \ll d$ coefficients is sufficient for the next layer. The protocol is detailed in Algorithm~\ref{alg:dcd}.

\begin{algorithm}[H]
\caption{Discrete Chebyshev Decomposition (DCD) Protocol}
\label{alg:dcd}
\KwIn{Output state of layer $k$, $\ket{\psi_{\text{out}}^{(k)}}$; truncation rank $r$; target precision $\delta$.}
\KwOut{Classical coefficient vector $\mathbf{c}_{\text{classical}} = [c_0, c_1, \dots, c_{r-1}]^T$.}
\BlankLine
\tcc{Stage 1: Quantum-to-Classical (Q2C) -- Coefficient Estimation}
\For{$j \leftarrow 0$ \KwTo $r-1$}{
    Efficiently prepare the basis state $\ket{T_j}$ using its known recurrence relation\;
    Construct a circuit to project $\ket{\psi_{\text{out}}^{(k)}}$ onto $\ket{T_j}$\;
    Use Quantum Amplitude Estimation (QAE) to estimate the coefficient $c_j = \braket{T_j | \psi_{\text{out}}^{(k)}}$ to precision $\delta$\;
    Store the estimated real value $c_j$ classically\;
}
\BlankLine
\tcc{Stage 2: Classical-to-Quantum (C2Q) -- State Re-preparation for Layer $k+1$}
Load the classical vector $\mathbf{c}_{\text{classical}}$ into QRAM\;
Use a QAM to compute the amplitudes of the approximate vector $\tilde{\psi}_i = \sum_{j=0}^{r-1} c_j T_{ji}$ for each computational basis state $\ket{i}$\;
Prepare the input state for layer $k+1$, $\ket{\psi_{\text{in}}^{(k+1)}} = \sum_i \tilde{\psi}_i \ket{i}$, using a standard state preparation routine.
\end{algorithm}

\begin{table*}[ht!]
\centering
\caption{Complexity Comparison of Information Extraction Protocols for a state in $\mathbb{C}^d$.}
\label{tab:complexity}
\begin{tabular}{@{}lccc@{}}
\toprule
\textbf{Protocol} & \textbf{Complexity} & \textbf{Classical Post-processing} & \textbf{Goal} \\ \midrule
\textbf{Full QST} & $O(d^2)$ & $O(d^3)$ & Full density matrix reconstruction \\
\textbf{Shadow Tomography}\textsuperscript{a} & $O(K \log(M)/\delta^2)$ & $O(M \cdot \text{poly}(\log d))$ & Estimate $M$ few-body observables \\
\textbf{DCD Protocol (Our work)} & $\boldsymbol{O(r / \delta)}$ & $\boldsymbol{O(r \cdot d)}$ & \textbf{Extract $r$ global feature coefficients} \\
\bottomrule
\end{tabular}
\\
\textsuperscript{a}{\footnotesize For estimating $M$ observables with Pauli weight at most $K$ to precision $\delta$.}
\end{table*}

\subsubsection{Complexity and Advantage}
The efficiency of the DCD protocol is summarized in Theorem~\ref{the-dcd}.
\begin{theorem}(Discrete Chebyshev Decomposition)
\label{the-dcd}
Given access to a state preparation unitary for $\ket{\psi}$ with cost $C_{\psi}$, the DCD protocol can estimate the first $r$ Chebyshev coefficients to precision $\delta$ with total query complexity $\tilde{O}(r \cdot C_{\psi} / \delta)$. The subsequent state re-preparation for the next layer requires a QRAM of size $r$ and a QAM with complexity $\tilde{O}(r \cdot \text{poly}(\log d))$.
\end{theorem}
It is worth noting that the DCD can be executed in parallel based on the quantum inner product estimation algorithm~\cite{xiong2024circuit}, to process one batch at a time.

As summarized in Table~\ref{tab:complexity}, the DCD protocol offers a clear advantage over the intractable scaling of QST. It also provides a powerful alternative to other modern techniques like shadow tomography. While shadows are highly effective for estimating local observables, DCD is purpose-built to extract a global, spectral representation of the state via its Chebyshev coefficients. The query complexity scales only with the desired number of features, $r$, and precision, $\delta$. Since our work demonstrates that $r$ can be significantly smaller than the Hilbert space dimension $d$, DCD transforms the data transfer from an insurmountable bottleneck into a manageable subroutine, making a true end-to-end quantum speedup for deep learning finally achievable.

\begin{figure*}[ht!]
    \centering
    \includegraphics[width=\linewidth]{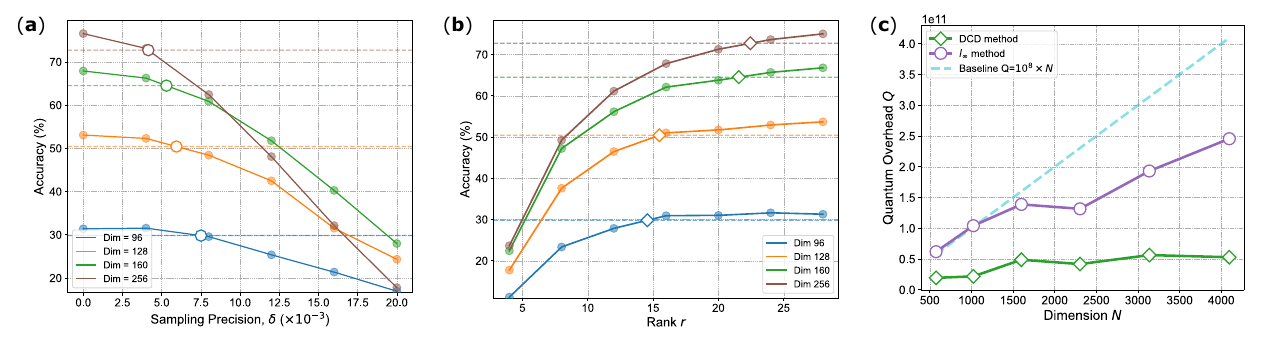}
    \caption{Comparative resource analysis of $l_{\infty}$ tomography and DCD on qResNet. 
    \textbf{(a, b)} Classification accuracy as a function of the respective hyperparameters (sampling precision for $l_{\infty}$ tomography, rank for DCD) across various input dimensions. The values required to achieve 95\% of peak performance are highlighted. 
    \textbf{(c)} Total quantum overhead $Q$ versus tensor dimension $N$ (where $N=(\text{input dimension}/4)^2$). Using the hyperparameters for 95\% peak performance, this plot compares the resource scaling of both methods against a $Q=N \times 10^8$ baseline. The results show that while the $l_{\infty}$ method has a slight sub-linear scaling, the DCD method demonstrates significantly lower resource consumption and a better scaling exponent.}
    \label{fig:hyperparameter_impact}
\end{figure*}

\section{Quantum Model Instances: qResNet \& qTransformer}
\label{sec:quantum_models}

To demonstrate the versatility of our framework, we now instantiate it by constructing a quantum Residual Network (qResNet) and a quantum Transformer (qTransformer). These examples showcase how our modular approach maps classical computational patterns onto the most suitable quantum primitives, guided by the principle of targeting speedups relative to the primary input dimension (e.g., sequence length $N$ or image size $H \times W$). The concrete quantum implementations of them can be found in Appendix~\ref{sec:q_res_block} and~\ref{sec-Attn-details}.

\subsection{Quantum ResNet (qResNet)}

Our Quantum ResNet (qResNet) adapts the architecture of a classical ResNet-18~\cite{he2016deep}. We focus on the regime where the image's spatial dimensions ($H, W$) are significantly larger than the channel dimension ($C$). The core computational tasks are Convolution, Activation, and Residual Connections, which are implemented using the QAM
A key design choice is the use of our Data Transfer Module (DTM) after each QAM-based layer. This Q2C-C2Q cycle prevents the composition of multiple sparse operators from creating a dense, computationally complex transformation, thereby preserving the efficiency of the QAM throughout the network's depth. The complexity is summarized in Theorem~\ref{the-qres}.

\begin{theorem}(Quantum ResNet Block)
\label{the-qres}
Given a ResNet Block, whose input tensor and kernel have shapes of $(B,C,H,W)$ and $(C,C,K,K)$ respectively, a quantum implementation of the ResNet Block has the quantum overhead of $\tilde{O}(CK^2\times S(B,C,H,W))$, where $S(B,C,H,W)$ is the sampling overhead of a quantum state with shape of $(B,C,H,W)$.
\end{theorem}

\subsection{Quantum Transformer (qTransformer)}

For vision tasks, we implement an encoder-only Quantum Transformer, focusing on the common $N \gg d$ regime, where $N$ is the sequence length and $d$ is the embedding dimension. This assumption dictates the allocation of tasks to create a highly efficient quantum analog, where Feed-Forward Network and Residuals are implemented solely by QAM. Multi-Head Self-Attention(MHSA) has a hybrid implementation of QLAM and QAM.
The complexity of the quantum MHSA module is given in Theorem~\ref{the-qtrans}
\begin{theorem}(Quantum Multi-Head Self-Attnetion Mechanism)
\label{the-qtrans}
For an input tensor with shape of $(B,N,d)$, where $N$ is the number of tokens and $d$ is the token length, the Quantum Overhead of the Quantum Multi-Head Self-Attnetion Mechanism is $O(d^2\log dB\textnormal{ polylog} N\times S(B,N,d)).$
\end{theorem}

The flowchart of the MHSA module (Figure~\ref{fig:framework_overview}(e)) showcases this strategic division of labor:
\begin{enumerate}
    \item \textbf{Input Projection (QAM):} The input embeddings $X \in \mathbb{R}^{N \times d}$ are projected to Query, Key, and Value representations by multiplying with small weight matrices ($W_Q, W_K, W_V \in \mathbb{R}^{d \times d}$). This operation can be viewed as $N$ parallel applications of a small linear map on $d$-dimensional vectors.
    \item \textbf{Attention Score Calculation (QAM):} The core classical bottleneck is computing the attention scores via the matrix product $S = QK^T$, which results in a large $N \times N$ matrix. This operation is a similar parallel computation.
    \item \textbf{Non-linear Activation (QAM):} An approximation to the softmax function is applied element-wise to the score matrix $S$ in parallel.
    \item \textbf{Output Calculation (QLAM):} The output is computed by multiplying the activated scores with the Value matrix, $S'V$. This is a dense matrix multiplication and is therefore executed by the QLAM. 
    \item \textbf{Aggregation Layer (QLAM):} The aggregation layer utilizes the Linear Combination of Unitaries (LCU) technique to naturally achieve an effective concatenation of the output states from each attention head.
\end{enumerate}
This deliberate allocation of computational tasks—reserving the QLAM for the true $N$-dimensional bottlenecks and using the QAM for structured, $d$-dimensional arithmetic—is the key to achieving a significant asymptotic speedup with respect to the input sequence length.

\begin{figure*}[ht!]
    \centering
    \includegraphics[width=\linewidth]{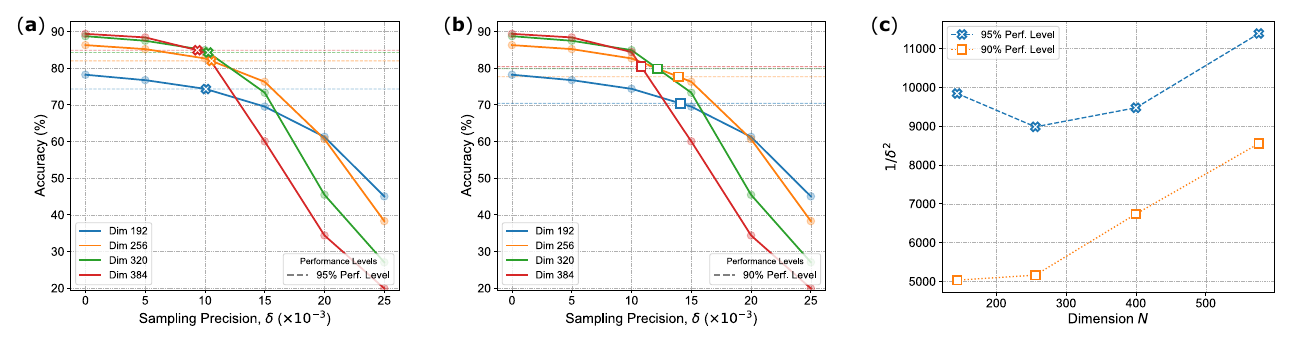}
    \caption{Performance and resource trade-off analysis of the DCD protocol on qResNet. 
    \textbf{(a, b)} Classification accuracy as a function of the DCD rank, with curves corresponding to achieving 90\% and 95\% of peak performance.
    \textbf{(c)} Quantum overhead $Q$ versus tensor dimension $N$. This plot directly compares the resource costs required to meet the 90\% and 95\% performance targets, illustrating the clear and predictable trade-off between accuracy and computational resources.}
    \label{fig:dcd_efficacy}
\end{figure*}

\section{Experiments}
\label{sec:experiments}

This section presents a series of numerical experiments designed to validate our proposed framework. The experimental validation is structured as follows. First, we conduct a direct comparative analysis of our Discrete Chebyshev Decomposition (DCD) method against a standard baseline, $l_{\infty}$ tomography, on a qResNet model. This experiment aims to establish the superior efficiency and scaling of DCD. Second, we perform an in-depth analysis of the DCD protocol's efficacy on both qResNet and qTransformer architectures. This investigation reveals distinct behavioral patterns and highlights the method's ability to exploit intrinsic model properties. Finally, we provide a concrete numerical resource analysis to quantify the advantages of our integrated approach. All experiments were conducted on standard image classification benchmarks, including CIFAR-10/100, CUB-200-2011, and Oxford-IIIT Pets.

\subsection{Comparative Analysis of Data Loading Methods on qResNet}
\label{sec:resnet_comparison}

To demonstrate the practical advantages of our proposed DCD method, we first benchmark it against the widely used $l_{\infty}$ tomography on a qResNet model. We evaluate how classification accuracy and resource overhead scale with the input dimension.

As depicted in Figure~\ref{fig:hyperparameter_impact}(a) and (b), we analyze the relationship between key hyperparameters and classification accuracy for varying input dimensions. For the $l_{\infty}$ tomography, the crucial hyperparameter is the sampling precision, while for our DCD method, it is the number of retained coefficients (rank). The plots illustrate the hyperparameter values required to achieve 95\% of the peak classification performance.

The primary findings are quantified in Figure~\ref{fig:hyperparameter_impact}(c), which plots the total quantum overhead, defined as $Q = (\text{quantum circuit T-depth}) \times (\text{sampling shots})$, against the effective tensor dimension $N$ (for ResNet, $N = (\text{input dimension} / 4)^2$). We plot the resource curves for both methods, using the hyperparameters identified for 95\% peak performance. For reference, a baseline of $Q=N \times 10^8$ is included. The results clearly indicate that while the $l_{\infty}$ tomography method achieves a favorable, slightly sub-linear scaling, our DCD method is significantly more efficient. DCD not only exhibits a substantially lower absolute resource cost but also demonstrates a more advantageous scaling exponent (i.e., a lower-degree polynomial scaling). This evidence strongly supports that DCD is a more scalable and resource-efficient data loading protocol for quantum convolutional architectures.

\begin{figure*}[ht!]
    \centering
    \includegraphics[width=\linewidth]{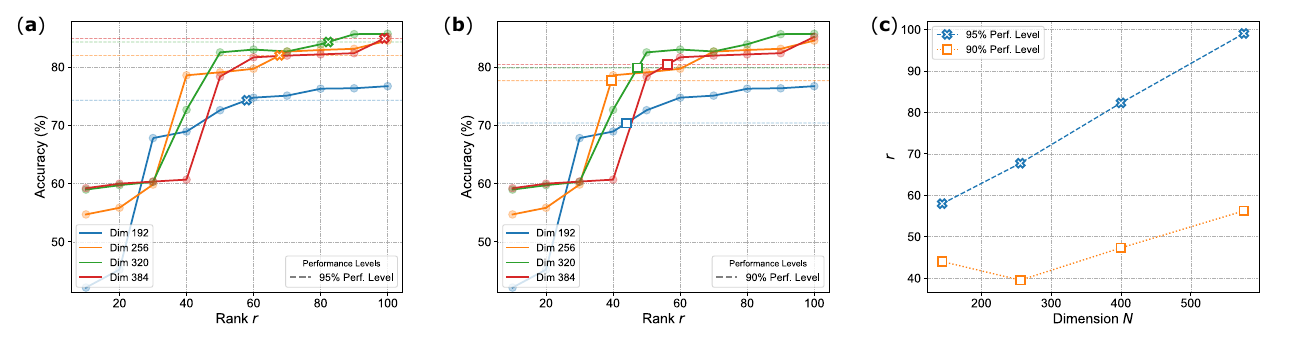}
    \caption{Analysis of the DCD protocol on qTransformer, revealing an intrinsic threshold effect.
    Across different input dimensions, the classification accuracy exhibits a sharp jump or ``elbow point'' as a function of the DCD rank. While the position of this threshold shifts with the input dimension, the phenomenon itself is stable. This suggests that DCD successfully isolates a compact, highly informative subspace that is critical to the qTransformer's performance.}
    \label{fig:qtrans_dcd}
\end{figure*}

\begin{figure}[ht!]
    \centering
    \includegraphics[width=0.8\linewidth]{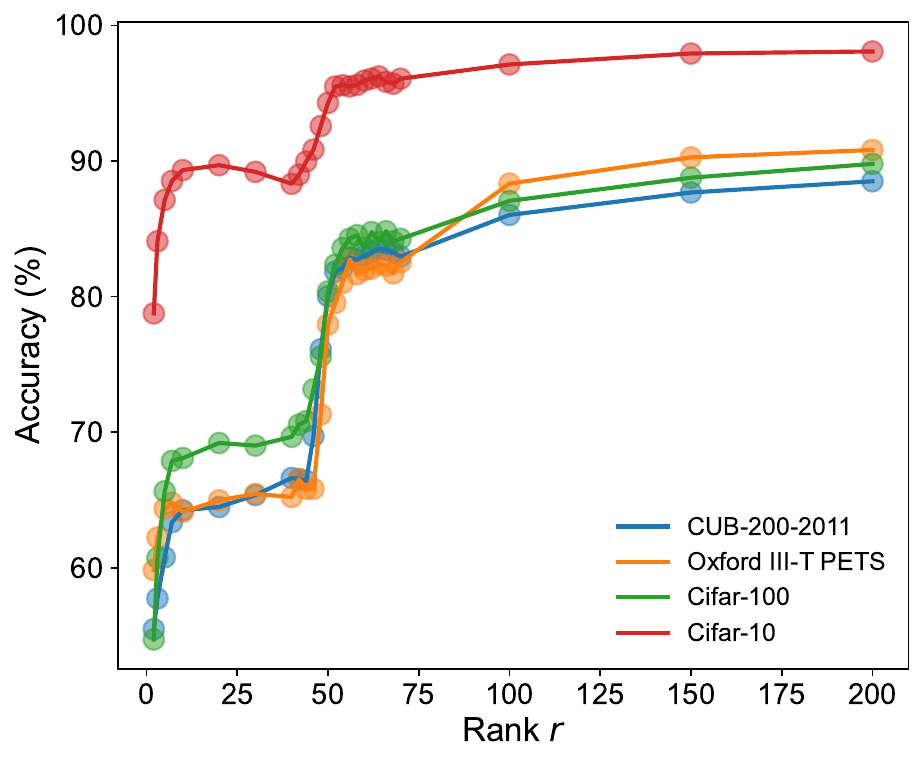}
    \caption{Validation of the DCD threshold effect across multiple datasets for the qTransformer. 
    The sharp increase in accuracy at a critical rank is consistently observed across different benchmark datasets. This provides strong evidence that this phenomenon is an intrinsic property of the qTransformer architecture itself, rather than an artifact of a specific dataset's characteristics.}
    \label{fig:trans_dataset}
\end{figure}

\subsection{In-depth Analysis of DCD Efficacy and Dimensionality Dependence}
\label{sec:dcd_analysis}

Having established DCD's superior performance compared to the baseline, we now delve deeper into its behavior across different model architectures, specifically qResNet and qTransformer. This analysis highlights the method's varying but consistently favorable impact on different models.

For the qResNet model, we analyze the trade-off between performance and resource cost. Figure~\ref{fig:dcd_efficacy}(a) and (b) illustrate the relationship between the DCD rank and the resulting classification accuracy, showing curves for achieving both 90\% and 95\% of peak performance. This data is then used in Figure~\ref{fig:dcd_efficacy}(c) to plot the quantum overhead $Q$ versus the tensor dimension $N$. By comparing the 90\% and 95\% performance curves, we observe a clear and predictable trade-off: a modest relaxation in the target accuracy translates into significant savings in quantum resources, making the cost-benefit analysis for practical applications straightforward.

The evaluation of DCD on the qTransformer architecture reveals a distinct and highly significant phenomenon. As shown in Figure~\ref{fig:qtrans_dcd}, the classification accuracy as a function of the DCD rank presents a sharp ``elbow'' point, or a sudden jump, at a specific rank value. This behavior is consistent across different input dimensions, although the position of the jump varies. This suggests a phase transition where a small set of coefficients contributes disproportionately to the model's performance.

Crucially, as demonstrated in Figure~\ref{fig:trans_dataset}, this threshold effect is not an artifact of a specific dataset but holds true across multiple benchmarks for the same model. This leads to a powerful conclusion: the remarkable importance of a small subset of decomposition coefficients is an intrinsic property of the qTransformer architecture itself. The DCD method is exceptionally well-suited to identify and isolate this compact, information-rich subspace within the Transformer's internal representations. By comparing the overall results, and considering the tensor dimension scaling for Transformer ($N = (\text{input dimension} / 16)^2$), it is evident that DCD's effectiveness on qTransformer shows an even lower dependency on input dimension, reinforcing its suitability for large-scale, complex quantum models.

\subsection{Numerical Resource Analysis}
\label{sec:numerical_analysis}

To ground our findings in practical terms, we provide a concrete resource analysis based on the execution of a single qResNet18 residual block, with results summarized in Table~\ref{tab:infidelity_overhead} and Table~\ref{tab:performance_overhead_tab}. This analysis validates our framework's advantages from two perspectives.

First, we demonstrate the inherent efficiency of our model architecture. We compare our architecture (using $l_{\infty}$-tomography for measurement) against a prior qCNN model~\cite{Kerenidis2020Quantum}. As shown in Table~\ref{tab:infidelity_overhead}(a), we measure the deviation ($l_2$-norm of the difference) between the output state of the quantum circuit and its ideal classical counterpart. Our model achieves a target fidelity with a significantly lower T-depth, confirming its superior computational efficiency over previous architectures.

Second, we compare the data loading protocols on our own architecture. Table~\ref{tab:performance_overhead_tab}(b) contrasts the end-to-end classification accuracy of our model when equipped with $l_{\infty}$-tomography versus our DCD method, under an equivalent resource budget. The results clearly show that the DCD-based model achieves substantially higher classification accuracy for a given cost. This highlights DCD's superior performance-to-cost ratio and better accuracy scaling, especially in resource-constrained scenarios. While $l_{\infty}$-tomography might achieve a marginally higher absolute peak accuracy given an exhaustive resource budget, DCD provides a much more practical and efficient pathway to achieving high performance, delivering the best results when resources are limited. This provides direct numerical evidence for the practical savings and superior performance of our fully integrated framework.

\begin{table*}[htbp]
    \centering
    \caption{Infidelity and Quantum Overhead for Different Models and Parameters. This table compares our model against the Full-QLA-Based approach proposed in ~\cite{Kerenidis2020Quantum} for a ResNet block. Infidelity is defined as the $l_2$-norm of the difference between the quantum circuit's output state and its ideal classical counterpart. The results show that to achieve a comparable level of infidelity, our model requires significantly lower quantum overhead.}
    \label{tab:infidelity_overhead}
    
    \small 
    \setlength{\tabcolsep}{3pt} 
    
    \begin{tabular}{c c cc cc cc cc} 
        \toprule
        \multirow{2}{*}{\textbf{Model}} & \multirow{2}{*}{\textbf{$M$}} & \multicolumn{8}{c}{\textbf{Sampling Precision}} \\ 
        \cmidrule(lr){3-10}
         & & \multicolumn{2}{c}{0.020} & \multicolumn{2}{c}{0.010} & \multicolumn{2}{c}{0.004} & \multicolumn{2}{c}{0.002} \\
        \cmidrule(lr){3-4} \cmidrule(lr){5-6} \cmidrule(lr){7-8} \cmidrule(lr){9-10}
         & & Inf. & QO & Inf. & QO & Inf. & QO & Inf. & QO \\
        \midrule
        
        Our Model & - 
        & $\bm{2.03 \!\times\! 10^{-2}}$ & $\bm{1.04 \!\times\! 10^{10}}$ 
        & $\bm{4.85 \!\times\! 10^{-3}}$ & $\bm{4.18 \!\times\! 10^{10}}$ 
        & $\bm{8.34 \!\times\! 10^{-4}}$ & $\bm{2.61 \!\times\! 10^{11}}$ 
        & $\bm{2.10 \!\times\! 10^{-4}}$ & $\bm{1.04 \!\times\! 10^{12}}$ \\
        \midrule
        
        \multirow{5}{*}{\makecell{Full QLA-Based \\ Model}} 
        & $10^{4}$ 
        & $2.26 \!\times\! 10^{-2}$ & $8.73 \!\times\! 10^{14}$ 
        & $9.00 \!\times\! 10^{-3}$ & $3.49 \!\times\! 10^{15}$ 
        & $5.75 \!\times\! 10^{-3}$ & $2.18 \!\times\! 10^{16}$ 
        & $5.51 \!\times\! 10^{-3}$ & $8.73 \!\times\! 10^{16}$ \\
        
         & $3 \!\times\! 10^{4}$ 
        & $1.96 \!\times\! 10^{-2}$ & $2.62 \!\times\! 10^{15}$ 
        & $5.10 \!\times\! 10^{-3}$ & $1.05 \!\times\! 10^{16}$ 
        & $1.18 \!\times\! 10^{-3}$ & $6.55 \!\times\! 10^{16}$ 
        & $4.91 \!\times\! 10^{-4}$ & $2.62 \!\times\! 10^{17}$ \\
        
         & $10^{5}$ 
        & $1.93 \!\times\! 10^{-2}$ & $8.73 \!\times\! 10^{15}$ 
        & $4.73 \!\times\! 10^{-3}$ & $3.49 \!\times\! 10^{16}$ 
        & $8.50 \!\times\! 10^{-4}$ & $2.18 \!\times\! 10^{17}$ 
        & $2.24 \!\times\! 10^{-4}$ & $8.73 \!\times\! 10^{17}$ \\
        
         & $3 \!\times\! 10^{5}$ 
        & $1.93 \!\times\! 10^{-2}$ & $2.62 \!\times\! 10^{16}$ 
        & $4.73 \!\times\! 10^{-3}$ & $1.05 \!\times\! 10^{17}$ 
        & $8.22 \!\times\! 10^{-4}$ & $6.54 \!\times\! 10^{17}$ 
        & $2.01 \!\times\! 10^{-4}$ & $2.62 \!\times! 10^{18}$ \\
        
         & $10^{6}$ 
        & $1.92 \!\times\! 10^{-2}$ & $8.73 \!\times\! 10^{16}$ 
        & $4.73 \!\times\! 10^{-3}$ & $3.49 \!\times\! 10^{17}$ 
        & $7.49 \!\times\! 10^{-4}$ & $2.18 \!\times\! 10^{18}$ 
        & $2.03 \!\times\! 10^{-4}$ & $8.73 \!\times\! 10^{18}$ \\
        \bottomrule
    \end{tabular}
    \vspace{1ex} 
    \footnotesize 
    \textit{Note: ``Inf.'' is an abbreviation for Infidelity, and ``QO'' is for Quantum Overhead.}
\end{table*}

\begin{table*}[htbp]
    \caption{
        Classification Accuracy and Quantum Overhead for Different Data Transfer Methods.
        This table compares the performance and resource cost of two protocols: our proposed Discrete Chebyshev Decomposition (DCD) and the baseline $l_{\infty}$ tomography.
        For a given target classification accuracy, the method that achieves it with lower quantum overhead (i.e., demonstrates a resource advantage) is highlighted in \textbf{bold}.
        The results clearly show that our DCD method excels in resource-constrained scenarios, showcasing its advantage by achieving high accuracy with significantly less computational cost.
    }
    \label{tab:performance_overhead_tab}
    \begin{threeparttable}
    \centering
    \setlength{\tabcolsep}{3.5pt} 
    \begin{tabular}{c c cc cc cc cc cc cc} 
        \toprule
        \multirow{2}{*}{\textbf{Method}} & \multirow{2}{*}{\textbf{Rank}} & \multicolumn{12}{c}{\textbf{Sampling Precision}} \\
        \cmidrule(lr){3-14}
         & & \multicolumn{2}{c}{0.0010} & \multicolumn{2}{c}{0.0020} & \multicolumn{2}{c}{0.0040} & \multicolumn{2}{c}{0.0100} & \multicolumn{2}{c}{0.0200} & \multicolumn{2}{c}{0.0400} \\
        \cmidrule(lr){3-4} \cmidrule(lr){5-6} \cmidrule(lr){7-8} \cmidrule(lr){9-10} \cmidrule(lr){11-12} \cmidrule(lr){13-14}
         & & Acc. & QO\tnote{a} & Acc. & QO\tnote{a} & Acc. & QO\tnote{a} & Acc. & QO\tnote{a} & Acc. & QO\tnote{a} & Acc. & QO\tnote{a} \\
        \midrule
        
        \multirow{5}{*}{\parbox{2cm}{\centering DCD}} 
        & 10 
        & 56.96 & 52.2 & 56.44 & 26.1 & 56.80 & 13.1 & 56.23 & 5.23 & \textbf{55.68} & \textbf{2.63} & \textbf{49.78} & \textbf{1.32} \\
        
         & 15 
        & 65.72 & 117 & 66.15 & 58.8 & 66.81 & 29.4 & 64.67 & 11.8 & \textbf{64.39} & \textbf{5.91} & 53.57 & 2.97 \\
        
         & 20 
        & 70.88 & 209 & 70.33 & 104 & 70.54 & 52.3 & 69.30 & 20.9 & \textbf{67.85} & \textbf{10.5} & 55.35 & 5.29 \\
        
         & 25 
        & 72.90 & 326 & 72.45 & 163 & 72.16 & 81.7 & \textbf{72.42} & \textbf{32.7} & \textbf{70.12} & \textbf{16.4} & 54.80 & 8.26 \\
        
         & 30 
        & 74.75 & 470 & 74.16 & 235 & \textbf{74.49} & \textbf{118} & \textbf{73.32} & \textbf{47.1} & 70.47 & 23.6 & 57.21 & 11.9 \\
        \midrule
        
        $l_{\infty}$ Tomo & - 
        & 75.23 & 3960 & \textbf{75.27} & \textbf{990} & 73.47 & 247 & 58.65 & 39.6 & 22.76 & 9.90 & 2.24 & 2.47 \\
        
        \bottomrule
    \end{tabular}
    
    \begin{tablenotes}
        \item[*] \footnotesize \textit{Note: ``Acc.'' is an abbreviation for Accuracy (\%), and ``QO'' is for Quantum Overhead.}
        \item[a] \footnotesize \textit{The values in this column are presented in units of $10^9$.}
    \end{tablenotes}

    \end{threeparttable}
\end{table*}

\section{Discussion}

Our work introduces a pragmatic and modular framework for constructing deep quantum networks, addressing the long-standing challenge of building multi-layer architectures. By strategically delegating tasks to specialized Quantum Arithmetic (QAM) and Quantum Linear Algebra (QLAM) modules, we can faithfully replicate complex models like ResNet and Transformer. The cornerstone of this approach is the Data Transfer Module (DTM), which enables this modularity. By intentionally forgoing a fully coherent quantum process between layers, we gain crucial architectural flexibility and the ability to implement strong non-linearities, albeit at the cost of intermediate measurements. Our proposed DCD protocol significantly mitigates this cost by efficiently compressing the inter-layer state information.

A pivotal assumption underpinning our framework's success is the inherent compressibility of the feature representations learned by these models. Our empirical results provide strong validation for this hypothesis. In particular, the sharp, phase-transition-like behavior observed in the qTransformer (Figure~\ref{fig:trans_dataset}) suggests that its internal states possess an intrinsic low-rank structure that is exceptionally well-captured by the DCD basis. This finding has profound implications, as efficient data loading and readout are the primary enablers for any end-to-end quantum speedup. It transforms the ``data loading problem'' from an insurmountable obstacle into a feature to be exploited. A crucial direction for future work is therefore to investigate the universality of this compressibility: is it a fundamental property of deep learning models, or is it task- and architecture-dependent? Understanding this could unlock further optimizations for a wide range of quantum machine learning algorithms.

However, the modularity of our framework introduces its own set of challenges and trade-offs. Each DTM call constitutes a full ``quantum-classical-quantum'' (QCQ) cycle, involving measurement and state re-preparation. While we have shown this cycle can be made highly efficient, it inevitably incurs overhead and potential information loss compared to a fully coherent computation. This opens up important avenues for future research. For instance, can we develop adaptive models that invoke the DTM only when necessary, such as when the accumulated transformation becomes too dense or non-linear effects are required? This could lead to hybrid models that balance the benefits of coherent evolution with the flexibility of our modular approach.

Finally, our analysis is situated within the paradigm of fault-tolerant quantum computing. While this allows us to establish valuable asymptotic scaling laws, translating these principles to near-term, noisy intermediate-scale quantum (NISQ) hardware remains a formidable challenge. Even for our most hardware-efficient qResNet variant, the circuit depth for a single block execution remains in the thousands, far beyond the reach of current devices. Therefore, a key future direction is to explore how our central design philosophy—modular construction and ``good-enough'' information transfer—can be adapted to the constraints and opportunities of the NISQ era. Could a ``NISQ-DCD'' protocol be developed that leverages variational circuits or error-mitigation techniques? Bridging this gap between fault-tolerant theory and near-term practice is essential for realizing the first practical quantum deep learning applications.

\begin{acknowledgements}
This work has been supported by the National Key Research and Development Program of China (Grant Nos. 2023YFB4502500 and 2024YFB4504100), the National Natural Science Foundation of China (Grant No. 12404564), and the Anhui Province Science and Technology Innovation (Grant Nos. 202423s06050001 and 202423r06050002).
\end{acknowledgements}

\bibliography{ref.bib}

\appendix



\section{Preliminaries\label{sec:preliminaries}}

To construct our framework for deep quantum networks, we leverage advanced algorithms from quantum linear algebra and quantum arithmetic, applying them to emulate classical architectures like ResNet and the Transformer. This section reviews these fundamental building blocks and contextualizes our work by detailing a critical bottleneck faced by existing approaches: the prohibitive cost of the quantum-classical data interface in multi-layer models.

\subsection{Quantum Subroutines for Linear Algebra}

Quantum Linear Algebra (QLA) promises significant speedups for classically intractable tasks, forming the computational core of many quantum machine learning proposals. While early algorithms like HHL demonstrated the potential for exponential advantage, modern QLA has largely converged around more versatile and robust techniques.

A central concept in modern QLA is block-encoding, a method for embedding a non-unitary matrix $A$ into a larger unitary matrix $U_A$. Specifically, an $(\alpha, a, \delta)$-block-encoding of $A$ is a unitary $U_A$ such that
\begin{equation}
    (\langle 0 |^{\otimes a} \otimes I) U_A (|0\rangle^{\otimes a} \otimes I) = A / \alpha,
\end{equation}
where $\alpha \ge \|A\|$ is a normalization factor, $a$ is the number of ancillary qubits, and the approximation is up to an error $\delta$. This technique transforms the problem of applying a matrix into the problem of implementing a unitary circuit, making it amenable to quantum computation. Many efficient block-encoding methods exist for structured matrices, such as sparse or low-rank matrices.

Once a matrix is block-encoded, its properties can be manipulated. For instance, the Quantum Singular Value Transformation (QSVT)~\cite{gilyen2019quantum} provides a unified framework for applying polynomial functions of a matrix's singular values to a quantum state. While QSVT is a powerful and general tool, many QLA tasks, including those in our work, can be realized using a more fundamental subroutine: Quantum Amplitude Estimation (QAE)~\cite{brassard2002quantum}. QAE allows for the estimation of the amplitude of a specific basis state in a quantum superposition. For example, if a quantum state $|\psi\rangle$ is prepared such that the probability of measuring a target state $|0\rangle$ is $p = |\langle 0 | \psi \rangle|^2$, QAE can estimate $p$ with an error $\delta$ using $O(1/\delta)$ queries to the state preparation circuit, achieving a quadratic speedup over classical sampling. This subroutine is crucial for extracting information from a quantum system, such as computing the inner product between two states or the expected value of an observable.

\subsection{Quantum Arithmetic for Sparse and Element-wise Operations}

While QLA excels at large-scale, dense matrix operations, DNNs also rely heavily on element-wise operations, such as adding biases, applying activation functions, and executing sparse transformations. These tasks necessitate Quantum Arithmetic (QA), which performs computations directly on the numerical values encoded in quantum registers, typically using a fixed-point binary representation.

QA circuits for fundamental operations like addition and multiplication have been well-established~\cite{vedral1996quantum, draper2000addition}, with resource costs (e.g., gate count and circuit depth) scaling polynomially with the precision (number of bits) of the encoded numbers. 
Furthermore, by adapting logic from classical circuits, quantum computers can efficiently perform general-purpose arithmetic operations.
\begin{lemma}\label{appen-quantum-arithmetic}
    Given a basic function \(f(x): \mathbb{R} \to \mathbb{R}\), there exists a quantum algorithm to implement quantum arithmetic \(|x\rangle|0\rangle \to |x\rangle|\tilde{f}(x)\rangle\), where \(|\tilde{f}(x) - f(x)| \leq \delta\) and \(\delta\) represents the computing accuracy. The gate complexity of the algorithm is \(O({\rm polylog}(1/\delta))\).
\end{lemma}
This capability is particularly vital for implementing sparse operations. A sparse matrix-vector multiplication, for instance, can be realized by arithmetically computing the non-zero matrix elements and adding the results to the corresponding components of the output vector. While this process is slower than the highly parallelized approach of QLA for dense matrices, its complexity scales with the number of non-zero elements, making it an efficient choice for sparse problems. Furthermore, QA is the primary tool for implementing non-linear activation functions, typically by computing a piecewise polynomial approximation of the target function (e.g., ReLU), which involves a sequence of arithmetic comparisons and calculations.

\subsection{Classical Architectures of Interest}

Our work focuses on developing quantum counterparts for two of the most influential DNN architectures.

The Transformer~\cite{vaswani2017attention} has become the de facto standard for sequence modeling tasks. Its core innovation is the self-attention mechanism, defined as $\text{Attention}(Q,K,V) = \text{softmax}(\frac{QK^T}{\sqrt{d_k}})V$. The primary computational bottleneck is the matrix multiplication $QK^T$, which scales as $O(N^2)$ with the sequence length $N$, making it a prime target for quantum acceleration via QLA.

The Residual Network (ResNet)~\cite{he2016deep} introduced the concept of residual connections, $y = \mathcal{F}(x) + x$, where $\mathcal{F}(x)$ is a block of layers. This ``shortcut'' structure effectively mitigates the vanishing gradient problem, enabling the training of networks with hundreds or even thousands of layers. Quantum analogues of ResNet provide an ideal testbed for assessing the ability of a quantum framework to handle truly deep architectures.


\section{Details of Quantum Arithmetics Implementation}
\label{sec:quantum_arithmetics}

\subsection{Tensor Representation}
\label{subsec:tensor}
In classical computing, a tensor is a multi-dimensional array of numbers. In the quantum arithmetic module, a tensor is represented by a composite quantum state, which is the tensor product of individual quantum registers. Let $T\in \mathbf{R}^{n_1\times n_2\times\cdots \times n_l}$ be a tensor. Its quantum representation $\ket{T}$ is encoded in a system of $N=\prod_i n_i$ quantum registers:
\begin{equation}
    \ket{T} = \bigotimes_{i_1,i_2,\dots,i_l} \ket{T_{i_1i_2\dots i_l}}.
\end{equation}
This state representation can also be identified with a quantum operation $O_T$ 
\begin{equation}
    O_T\ket{0}= \ket{T}.
\end{equation}
A controlled state preparation can support parallel computation, such as for $O_T^{(c)}$:
\begin{equation}
    O_T^{(c)}\ket{i_1,i_2,\cdots,i_{p}}\ket{0}= \ket{i_1,i_2,\cdots,i_{p}}\bigotimes_{i_{p+1},i_{p+2},\dots,i_l} \ket{T_{i_1i_2\dots i_l}},
\end{equation}
an addition on the first register implies parallel addition operations for every element that $i_{p+1},i_{p+2},\dots,i_{l}=0$.

\subsection{Basic Operations}
\label{subsec:basic_ops}
Basic arithmetic operations like addition and multiplication form the building blocks for more complex functions. These are implemented as reversible quantum circuits.



\paragraph{Go Forward.}
The quantum implementation of the tensor operation, such as tensor addition and dot, can be inherited from classical computing, except the uncomputation process, since the invertibility of the quantum operations. 

\paragraph{Uncomputation.}
To maintain the reversibility and release ancillary qubits for reuse, an uncomputation step is crucial. For instance, after computing $\ket{a}\ket{a+b}$, one might need to restore the second register to its original state $\ket{b}$. This is achieved by applying the inverse of the adder circuit, $U_{\text{add}}^\dagger$.

\paragraph{Tensor Dot.}
Take the tensor dot as an example, we explain how a typical arithmetic process on a quantum computer. Given the two tensors $S\in\mathbb{R}^{c_s\times d\times p}, T\in\mathbb{R}^{c_t\times d\times q}$, the output of the tensor dot is defined by 
$$R_{ijkl}=\sum_{\mu}S_{i\mu j}T_{k\mu l}.$$
Suppose the quantum representations is $O_T^{(c)}$, whose functional is
$$
O_T^{(c)}\ket{k}\ket{0^{\otimes d}}\ket{0^{\otimes q}}=|k\rangle\otimes_{\mu,l}\ket{T_{k\mu l}},
$$
similarly for $O_S^{(c)}$. The dot operation is implemented as below:
\begin{equation}
    \begin{split}
        &O_T^{(c)}\ket{i}\ket{0^{\otimes d+p}}\otimes O_S^{(c)}\ket{k}\ket{0^{\otimes d+q}}\otimes\ket{0}\\
        =&\ket{i,k}\otimes_{j,l}(\otimes_{\mu}\ket{S_{i\mu j}, T_{k\mu l}})\otimes\ket{0}\\
        \to&\ket{i,k}\otimes_{j,l}(\otimes_{\mu}\ket{S_{i\mu j}, T_{k\mu l}})\otimes\ket{R_{ijkl}}
    \end{split}
\end{equation}

\subsection{Element-wise operations}
The design of element-wise operations can directly follow classical computing. Special circuits exist to optimize the overhead for specific operations, given the difference between quantum and classical ALUs.

\paragraph{\textbf{Reciprocal}}

We use the Newton method to calculate the reciprocal on a quantum computer \cite{bhaskar2015quantum}. This target is expressed as:
\begin{equation}
    |x,0\rangle\to|x,\frac{1}{x}\rangle.
\end{equation}
This can be approximately achieved through the following iteration:
\begin{equation}
    x_{(k+1)}=x_{(k)}(2-xx_{(k)}).
\end{equation}

\paragraph{\textbf{Arc cosine}}
The Quantum Function-value Binary Expansion method is chosen to calculate $\arccos$ \cite{Wang2020}, which realizes approximately the transformation
\begin{equation}
    |x,0\rangle\to|x,\arccos{x}\rangle.
\end{equation}
The iteration reads
\begin{equation}
    x_{(0)} = x, \quad x_{(k+1)}=
    \begin{cases}
        2x_{(k)}^2-1,\quad &x_{(k)}>0,\\
        1-2x_{(k)}^2,\quad &x_{(k)}\leq0.
    \end{cases}
\end{equation}

\paragraph{\textbf{ReLU Activation}}
\label{subsec:relu}
The Rectified Linear Unit (ReLU), defined as $f(x) = \max(0, x)$, is a non-linear activation function. Since it is not inherently reversible (e.g., both -2 and -5 map to 0), its quantum implementation must store the output in a separate register.
\begin{equation}
    U_{\text{ReLU}} \ket{x}\ket{0} = \ket{x}\ket{\max(0, x)}.
\end{equation}
The implementation leverages the fixed-point representation of $x$, where the most significant bit (MSB) serves as the sign bit (e.g., 0 for positive, 1 for negative in two's complement). The circuit operates as follows:
\begin{enumerate}
    \item An ancilla qubit is prepared to store the sign of $x$. This is typically the MSB of the $\ket{x}$ register.
    \item A controlled-copy operation is performed. The sign qubit acts as the control.
    \item If the sign bit is 0 (i.e., $x \ge 0$), the content of register $\ket{x}$ is copied to the output register. This can be achieved with a cascade of CNOT gates.
    \item If the sign bit is 1 (i.e., $x < 0$), no operation is performed, leaving the output register in its initial state $\ket{0}$.
\end{enumerate}
This circuit effectively implements a quantum ``if-then'' statement, enabling the non-linearity while preserving unitarity.

\section{Implementation details of a Residual Layer}

\label{sec:q_res_block}

To construct deep quantum neural networks, we introduce a quantum analogue of the classical residual block, inspired by ResNet architectures. This block enables the training of deeper models by using shortcut connections to mitigate vanishing gradient problems. A single block operates on a quantum state encoding a feature map and is composed of a main path and a shortcut path. The data flow within the block is managed by a Data Transfer Module (DTM), which handles state preparation from classical data via QRAM and measurement for intermediate classical processing.

A typical quantum residual block executes the following sequence:

1.  \textbf{Main Path:} The input state $\ket{\psi_{\text{in}}}$, encoding the feature map $X$, is processed sequentially by a quantum convolutional layer ($U_{\text{qConv}}$), a quantum batch normalization layer ($U_{\text{qBN}}$), and a quantum ReLU activation ($U_{\text{QReLU}}$). This sequence may be repeated, as in standard ResNet blocks.

2.  \textbf{Shortcut Path:} The original input state $\ket{\psi_{\text{in}}}$ is preserved.

3.  \textbf{Addition \& Final Activation:} The output state from the main path, $\ket{\psi_{\text{main}}}$, is added to the shortcut state $\ket{\psi_{\text{in}}}$ using a quantum arithmetic adder from the QAM. A final ReLU activation, $U_{\text{QReLU}}$, is applied to the resulting state $\ket{\psi_{\text{main}}} + \ket{\psi_{\text{in}}}$ to produce the block's output state, $\ket{\psi_{\text{out}}}$.

\algdef{SE}[SUBALG]{Indent}{EndIndent}{\algorithmicindent}{}%
\algtext*{Indent}
\algtext*{EndIndent}

\begin{algorithm}[ht]
\caption{Quantum Residual Block (Comments on New Lines)}
\label{alg:q_res_block_standard}
\begin{algorithmic}[1]
\Require 
    Input quantum state $\ket{\psi_{X}}$; Classical kernel $K$; Parameters $\gamma, \beta$.
\Ensure 
    Output quantum state $\ket{\psi_{\text{out}}}$.

\Statex \textbf{--- Main Path ---}
\State $\ket{\psi_{\text{conv}}} \leftarrow U_{\text{qConv}}(K) \ket{\psi_{X}}$ 
\CommentLine{Apply 3x3 Quantum Convolution.}

\Statex \textit{Hybrid Batch Normalization Step}
\State $X_{\text{conv}} \leftarrow \text{Measure}(\ket{\psi_{\text{conv}}})$ 
\CommentLine{DTM: Quantum-to-Classical.}

\State $\mu, \sigma^2 \leftarrow \text{ComputeBatchStats}(X_{\text{conv}})$ 
\CommentLine{Classical computation of stats.}

\State $\ket{\psi_{\text{BN}}} \leftarrow U_{\text{qBN}}(\mu, \sigma^2, \gamma, \beta) \ket{\psi_{\text{conv}}}$ 
\CommentLine{Apply Quantum Batch Norm.}

\State $\ket{\psi_{\text{main}}} \leftarrow U_{\text{QReLU}} \ket{\psi_{\text{BN}}}$ 
\CommentLine{Apply Quantum ReLU.}

\Statex \textbf{--- Shortcut and Addition ---}
\State $\ket{\psi_{\text{shortcut}}} \leftarrow \ket{\psi_X}$ 
\CommentLine{Identity shortcut path.}

\State $\ket{\psi_{\text{sum}}} \leftarrow \text{QuantumAdd}(\ket{\psi_{\text{main}}}, \ket{\psi_{\text{shortcut}}})$ 
\CommentLine{Element-wise addition.}

\Statex \textbf{--- Final Activation ---}
\State $\ket{\psi_{\text{out}}} \leftarrow U_{\text{QReLU}} \ket{\psi_{\text{sum}}}$ 
\CommentLine{Apply final Quantum ReLU.}

\State \textbf{return} $\ket{\psi_{\text{out}}}$
\end{algorithmic}
\end{algorithm}

\paragraph{Quantum Convolutional Layer (qConv).}
The qConv layer performs convolution using the Quantum Arithmetic Module (QAM). Its goal is to transform an input feature map state $\ket{\psi_X}$ into an output state $\ket{\psi_Y}$ where $Y$ is the convolution of $X$ with a classically-defined kernel $K$. The operation can be described as follows: for each output pixel position $(i, j, c)$, the QAM applies a unitary $U_{\text{qConv}}$ that computes the dot product arithmetically. This unitary acts on a target register, initialized to zero, transforming it based on the input data accessible via a QRAM-like mechanism:
\begin{equation}
    U_{\text{qConv}}: \ket{i,j,c}\ket{0} \rightarrow \ket{i,j,c}\ket{Y_{ijc}},
\end{equation}
where $Y_{ijc} = (\text{bias})_c + \sum_{c', \Delta i, \Delta j} K_{c, c', \Delta i, \Delta j} \cdot X_{i+\Delta i, j+\Delta j, c'}$. This computation leverages the quantum adders and multipliers within the QAM to perform the operation in superposition across all output positions.

\paragraph{Quantum Batch Normalization Layer (qBatchNorm).}
The qBatchNorm layer, crucial for stabilizing training, is implemented in a hybrid quantum-classical manner. Due to the difficulty of computing global statistics (mean and variance) on a quantum state directly, we first perform a measurement on the state produced by the qConv layer. 
This DTM operation yields a classical snapshot of the feature map data. From this classical data, we compute the batch mean $\mu$ and variance $\sigma^2$. These classical parameters are then used to configure a quantum arithmetic circuit $U_{\text{qBN}}$ within the QAM. This circuit applies the normalization transformation element-wise in superposition:
\begin{equation}
    U_{\text{qBN}}: \ket{y} \rightarrow \ket{\gamma \frac{y - \mu}{\sqrt{\sigma^2 + \delta}} + \beta},
\end{equation}
where $\gamma$ and $\beta$ are learnable classical parameters, and $\ket{y}$ is a register digitally encoding a single feature value. The subsequent Quantum ReLU ($U_{\text{QReLU}}$) is similarly implemented as an arithmetic comparison circuit within the QAM, applying $\ket{y} \rightarrow \ket{\max(0, y)}$.

\section{Implementation Details of the Quantum Transformer}
\label{sec-Attn-details}

This section details the quantum implementation of the Transformer architecture, which, like the quantum ResNet, is constructed from modular components. It leverages the Quantum Arithmetic Module (QAM) and the Quantum Linear Algebra Module (QLAM). We assume the input is a quantum state encoding a tensor of shape $N \times d$, where $N$ is the sequence length and $d$ is the embedding dimension. For long-context tasks where $N \gg d$, the key to efficiency lies in how these modules handle the different dimensions: the large dimension $N$ is parallelized over using index registers, while the smaller dimension $d$ is processed arithmetically.

\begin{algorithm}[ht]
\caption{Quantum Residual Block}
\label{alg:q_res_block_standard}
\begin{algorithmic}[1]
\Require 
    Input quantum state $\ket{\psi_{X}}$; Classical kernel $K$; Parameters $\gamma, \beta$.
\Ensure 
    Output quantum state $\ket{\psi_{\text{out}}}$.

\Statex \textbf{--- Main Path ---}
\State $\ket{\psi_{\text{conv}}} \leftarrow U_{\text{qConv}}(K) \ket{\psi_{X}}$
\CommentLine{Apply 3x3 Quantum Convolution.}

\Statex \textit{Hybrid Batch Normalization Step}
\State $X_{\text{conv}} \leftarrow \text{Measure}(\ket{\psi_{\text{conv}}})$
\CommentLine{DTM: Quantum-to-Classical.}

\State $\mu, \sigma^2 \leftarrow \text{ComputeBatchStats}(X_{\text{conv}})$
\CommentLine{Classical computation of stats.}

\State $\ket{\psi_{\text{BN}}} \leftarrow U_{\text{qBN}}(\mu, \sigma^2, \gamma, \beta) \ket{\psi_{\text{conv}}}$
\CommentLine{Apply Quantum Batch Norm.}

\State $\ket{\psi_{\text{main}}} \leftarrow U_{\text{QReLU}} \ket{\psi_{\text{BN}}}$
\CommentLine{Apply Quantum ReLU.}

\Statex \textbf{--- Shortcut and Addition ---}
\State $\ket{\psi_{\text{shortcut}}} \leftarrow \ket{\psi_X}$
\CommentLine{Identity shortcut path.}

\State $\ket{\psi_{\text{sum}}} \leftarrow \text{QuantumAdd}(\ket{\psi_{\text{main}}}, \ket{\psi_{\text{shortcut}}})$
\CommentLine{Element-wise addition.}

\Statex \textbf{--- Final Activation ---}
\State $\ket{\psi_{\text{out}}} \leftarrow U_{\text{QReLU}} \ket{\psi_{\text{sum}}}$
\CommentLine{Apply final Quantum ReLU.}

\State \textbf{return} $\ket{\psi_{\text{out}}}$
\end{algorithmic}
\end{algorithm}









\subsection{Quantum Multi-Head Self-Attention}
\label{ssec:q_attention}

The self-attention mechanism is a hybrid of QAM-based arithmetic for local operations and QLAM-based matrix multiplication for the final aggregation. The parallelism over the sequence length $N$ is achieved by encoding the token indices into dedicated quantum registers, allowing the QAM to operate on all elements in superposition.

\paragraph{Q, K, V Projection and Score Calculation (via QAM).}
The initial step projects the input state $\ket{\psi_X}$ into Query ($\bm{Q}$), Key ($\bm{K}$), and Value ($\bm{V}$) representations. This is not a single large matrix multiplication, but rather $N$ independent small ones (on the $d$-dimensional vectors). This is performed by the QAM, conditioned on an index register $\ket{i}$ spanning the $N$ tokens.

The subsequent calculation of the attention scores, $\bm{S} = \bm{Q}\bm{K}^T / \sqrt{d_k}$, which results in an $N \times N$ matrix, perfectly illustrates this principle. To compute all $N^2$ scores in parallel, we use two index registers, $\ket{i}$ and $\ket{j}$. An arithmetic circuit within the QAM then executes the dot product conditioned on these indices. The transformation on the quantum state can be abstractly represented as:
\begin{equation}
\label{eq:qam_dot_product}
U_{\text{dot-prod}}: \ket{i,j}\ket{Q_i}\ket{K_j}\ket{0} \rightarrow \ket{i,j}\ket{Q_i}\ket{K_j}\ket{S_{ij}}.
\end{equation}
Here, the state $\ket{i,j}$ acts as a control, specifying which dot product to compute, while the operation itself happens on the data registers. The states $\ket{Q_i}$ and $\ket{K_j}$ represent the necessary data for the computation, made accessible via a QRAM-like mechanism. This explicitly shows how the large $N \times N$ dimensional workload is handled through quantum parallelism rather than matrix size.

\paragraph{Hybrid Softmax and Final Aggregation (via QLAM).}
A full quantum implementation of the softmax function is notoriously difficult. We therefore adopt a hybrid quantum-classical approach. The state encoding the unnormalized score matrix $\bm{S}$ (as constructed in Eq.~\ref{eq:qam_dot_product}) is measured using the DTM. The $N \times N$ matrix of scores is then post-processed classically to compute the final attention matrix $\bm{A} = \text{softmax}(\bm{S})$.

The final and most computationally intensive step is the product $\bm{A}\bm{V}$. Here, $\bm{A}$ is a large, dense $N \times N$ matrix. Given our assumption that $N \gg d$, this large matrix multiplication is precisely the task for which the QLAM is designed. The classical matrix $\bm{A}$ is used to construct its block-encoding unitary, $U_A$. The QLAM then efficiently applies this unitary to the quantum state encoding the $\bm{V}$ matrix. This strategic division of labor—using QAM for index-based parallelism on local data and QLAM for the large global aggregation—is key to our model's efficiency.

\paragraph{Multi-Head Parallelism.}
The multi-head mechanism is implemented by introducing an additional ``head index'' register. The entire single-head attention process described above is executed in superposition, conditioned on the state of this head register. The resulting states from all heads are then concatenated and passed through a final linear projection ($\bm{W}^O$), which, being a small operation on the $d$-dimension, is again handled by the QAM.

\subsection{Quantum Feed-Forward Network (FFN)}
\label{ssec:q_ffn}

Each Transformer block contains a position-wise Feed-Forward Network (FFN), applied independently to each of the $N$ token positions. This sub-layer is implemented entirely using the QAM. The mechanism is identical to that in the attention layer: the FFN's arithmetic circuits (two linear maps and a QReLU) are conditioned on an index register $\ket{i}$ that spans all $N$ token positions, thus processing all tokens in parallel. A complete quantum Transformer block is then formed by enclosing both the multi-head attention and FFN modules within residual connections and layer normalization, which are also implemented as QAM-based arithmetic operations conditioned on the token index.

\section{Quantum-Accelerated Backpropagation}
\label{sec:q_backprop}

Training deep neural networks relies on backpropagation, which systematically computes the gradient of the loss function with respect to the model's weights. We propose a quantum-accelerated approach for this process, where the core matrix operations of the chain rule are mapped to our QAM and QLAM modules. The overall process remains hybrid: gradients are typically stored and updated classically, but their computationally expensive calculation is offloaded to the quantum processor.

To illustrate the principle, we consider the backward pass through a single linear layer, defined by the forward pass $\bm{Y} = \bm{W}\bm{X}$. Here, $\bm{W}$ is a $d \times d$ weight matrix and $\bm{X}$ is a $d \times N$ matrix representing $N$ data points. Given the gradient from the subsequent layer, $\partial L / \partial \bm{Y}$ (a $d \times N$ matrix), we must compute two quantities: the gradient to be propagated backward, $\partial L / \partial \bm{X}$, and the gradient for updating the weights, $\partial L / \partial \bm{W}$.

\paragraph{\texorpdfstring{Gradient Calculation for Weights ($\partial L / \partial \bm{W}$): Handled by QLAM.}{Gradient Calculation for Weights (dL/dW): Handled by QLAM.}}

The gradient with respect to the input is given by the chain rule:
\begin{equation}
    \frac{\partial L}{\partial \bm{X}} = \bm{W}^T \frac{\partial L}{\partial \bm{Y}}.
\end{equation}
This is a $(d \times d) \times (d \times N)$ matrix multiplication. Critically, this operation can be viewed as applying the small ($d \times d$) matrix $\bm{W}^T$ to each of the $N$ columns of the incoming gradient $\partial L / \partial \bm{Y}$. This is a ``position-wise'' operation, perfectly suited for the QAM. By conditioning on an index register $\ket{j}$ spanning the $N$ columns, the QAM can perform all $N$ matrix-vector products in parallel, arithmetically processing the $d$-dimensional vectors in superposition.

\paragraph{\texorpdfstring{Gradient Calculation for Weights ($\partial L / \partial \bm{W}$): Handled by QLAM.}{Gradient Calculation for Weights (dL/dW): Handled by QLAM.}}

The gradient with respect to the weights is an outer product:
\begin{equation}
    \frac{\partial L}{\partial \bm{W}} = \frac{\partial L}{\partial \bm{Y}} \bm{X}^T.
\end{equation}
This is a $(d \times N) \times (N \times d)$ matrix multiplication, resulting in a $d \times d$ gradient matrix. Each element $(\partial L / \partial \bm{W})_{ij}$ is the inner product of the $i$-th row of $\partial L / \partial \bm{Y}$ and the $j$-th row of $\bm{X}$. Both are vectors of length $N$. Given our assumption that $N \gg d$, these are high-dimensional inner products. This task is ideal for the QLAM's inner product estimation capability \cite{xiong2024circuit}. Instead of performing a full matrix multiplication, the QLAM can be configured to efficiently estimate the $d^2$ required inner products between the corresponding $N$-dimensional quantum states, yielding the elements of the weight gradient.

This strategic division of labor is fundamental to our training approach. The QAM handles the backward flow of gradients through the network's data path by parallelizing over the sequence/batch dimension $N$. The QLAM, in turn, handles the most intensive gradient calculations for weights, which involve contractions over this large $N$ dimension. This transforms the most demanding parts of backpropagation into potentially tractable quantum computations, paving the way for end-to-end quantum-accelerated training.

\begin{figure}[ht]
    \centering
    \includegraphics[width=\columnwidth]{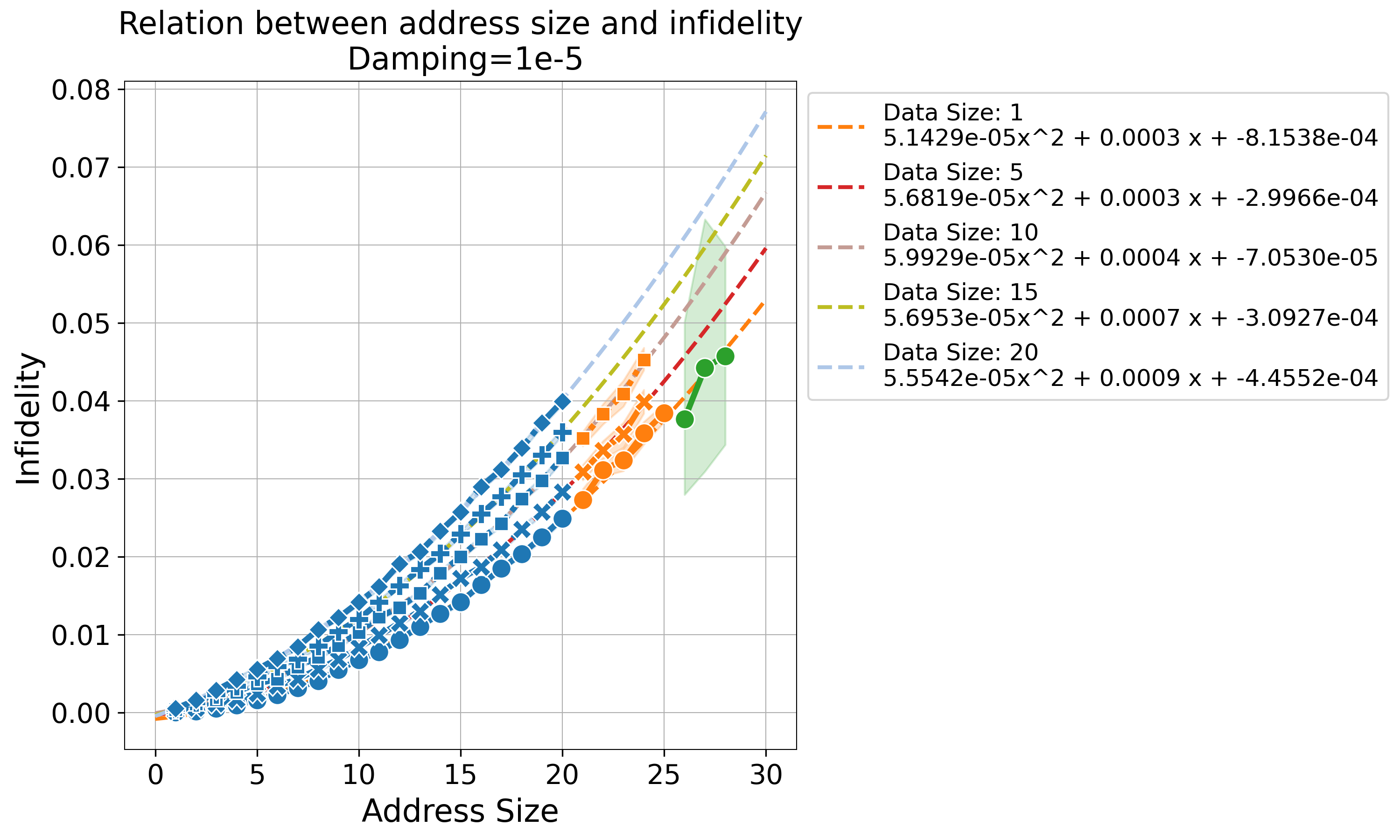}
    \caption{The figure shows the relation between address size and infidelity with the fixed k. The blue points are from the numerical experiments with 100 repetitions and each repetition of experiments consists of 1000 shots. The orange points are experiments with 10 repetitions with 1000 shots each. The Green ones are experiments with 10 repetitions with 10 shots each. The fitting function is completely decided by the data of blue points. }
    \label{fig:Fitting_Address_30_damping}
\end{figure}

\begin{figure}[ht]
    \centering
    \includegraphics[width=\columnwidth]{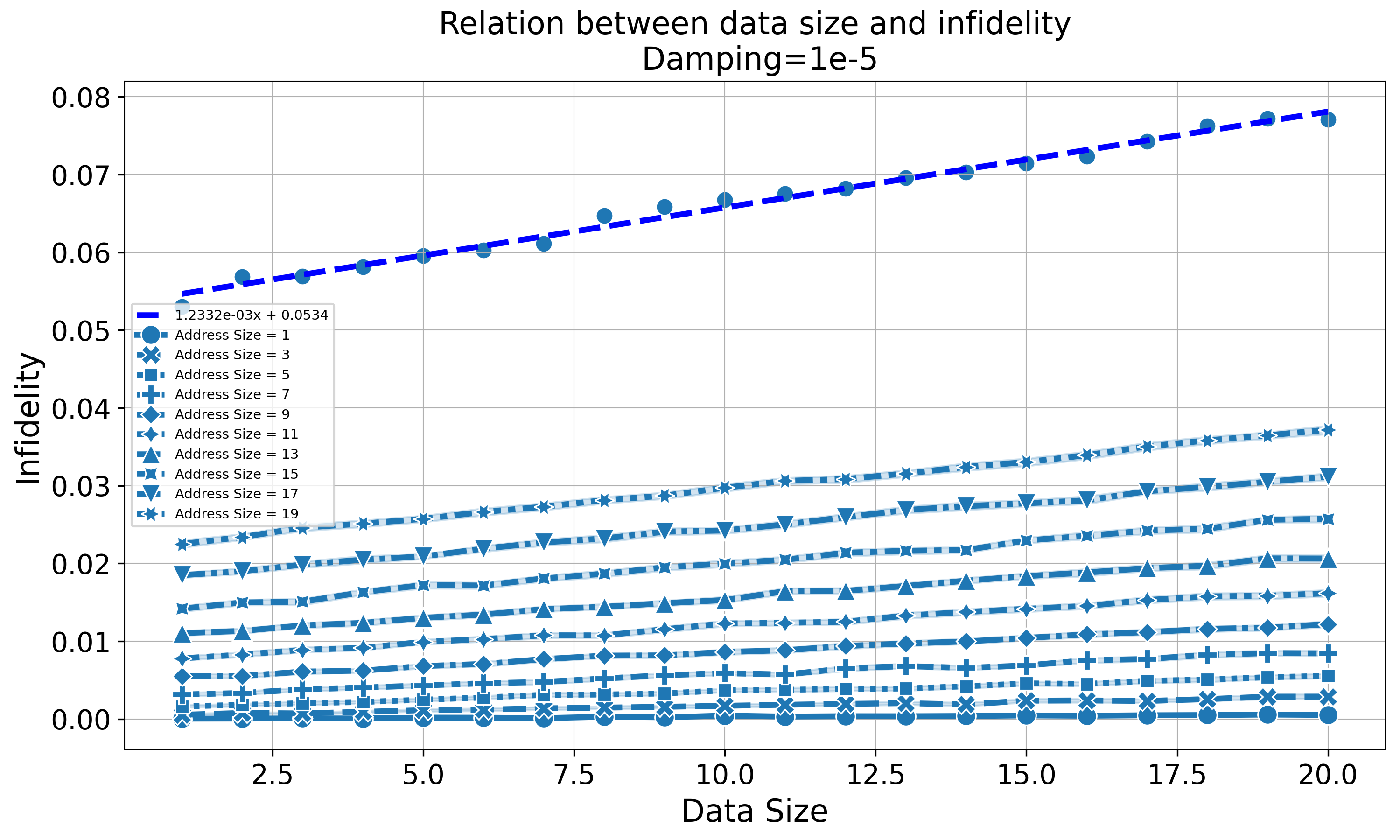}
    \caption{The figure presents the scattering points and sketches the linear relations between data size $k$ and the infidelity.}
    \label{fig:Fitting_data_64_damping}
\end{figure}

\section{Practicality of qRAM used in QViT\label{appen-qram}}

In this section, we introduce quantum random access memory (qRAM) \cite{giovannetti2008quantum}, a quantum architecture fundamental to our framework. QRAM serves as a generalization of classical RAM, leveraging quantum mechanical properties to enhance computational efficiency. 

In classical RAM, a discrete address \(i\) is provided as input, retrieving the memory element \(x_i\) stored at that location. Conversely, in qRAM, a quantum superposition of different addresses \(|\psi_{\textrm{in}}\rangle\) is input, and qRAM returns an entangled state \(|\psi_{\textrm{out}}\rangle\) where each address is correlated with the corresponding memory element:
\begin{equation}
    |\psi_{\textrm{in}}\rangle = \sum_{i=0}^{N-1} \alpha_i |i\rangle_A |0\rangle_D \xrightarrow{\textrm{qRAM}} |\psi_{\textrm{out}}\rangle = \sum_{i=0}^{N-1} \alpha_i |i\rangle_A |x_i\rangle_D,
\end{equation}
where \(N\) is the size of the data vector \(x\), and the superscripts \(A\) and \(D\) denote ``address'' and ``data'' respectively.

While we have characterized our QViT as the quantum deep learning framework in the fault-tolerant era, it is still imperative to incorporate the simulation of noisy qRAM. It is frequently used in QSave and QLoad in QViT. Within this framework, the primary role of qRAM is to retrieve pixel information stored in a massive matrix of size $2^{20}$ by $2^{10}$. Each pixel can hold either 32 or 64 bits, necessitating a $(30, 64)$ or $(30, 32)$-qRAM configuration. Our numerical simulations demonstrate promising results. For a $(30, 64)$-qRAM configuration, we observe an average state fidelity of $87\%$. This fidelity increases to $91\%$ for the $(30, 32)$-qRAM configuration.

The practicality of qRAM has been investigated on such a scale under our numerical experiments. Prior research indicates that qRAM infidelity scales as $O(n(n+k))$, where $n$ represents the address size and $k$ denotes the word length. This implies that infidelity exhibits quadratic growth with respect to address size for a fixed $k$ and increases with word length for a fixed address size $n$. Based on these established relationships, our experiments employed data with a fixed word length $k$ to maintain consistency with the established infidelity relation. Subsequently, we extrapolated the findings to the case of $n = 30$.  Using these extrapolated data $F(30, k)$, we employed a linear function to predict the infidelity value when $k = 64$.

All simulations were conducted under a controlled environment with $10^{-5}$ damping noise. We successfully simulated qRAM configurations ranging from $(20, 20)$ and below. The observed infidelities agree to the $O(n(n+k))$ relationship, demonstrating a quadratic dependence on address size $n$ for a fixed word length $k$, as shown in Figure~\ref{fig:Fitting_Address_30_damping} and Figure~\ref{fig:Fitting_data_64_damping}.

\begin{figure}[t]
    \centering
    \includegraphics[width=\columnwidth]{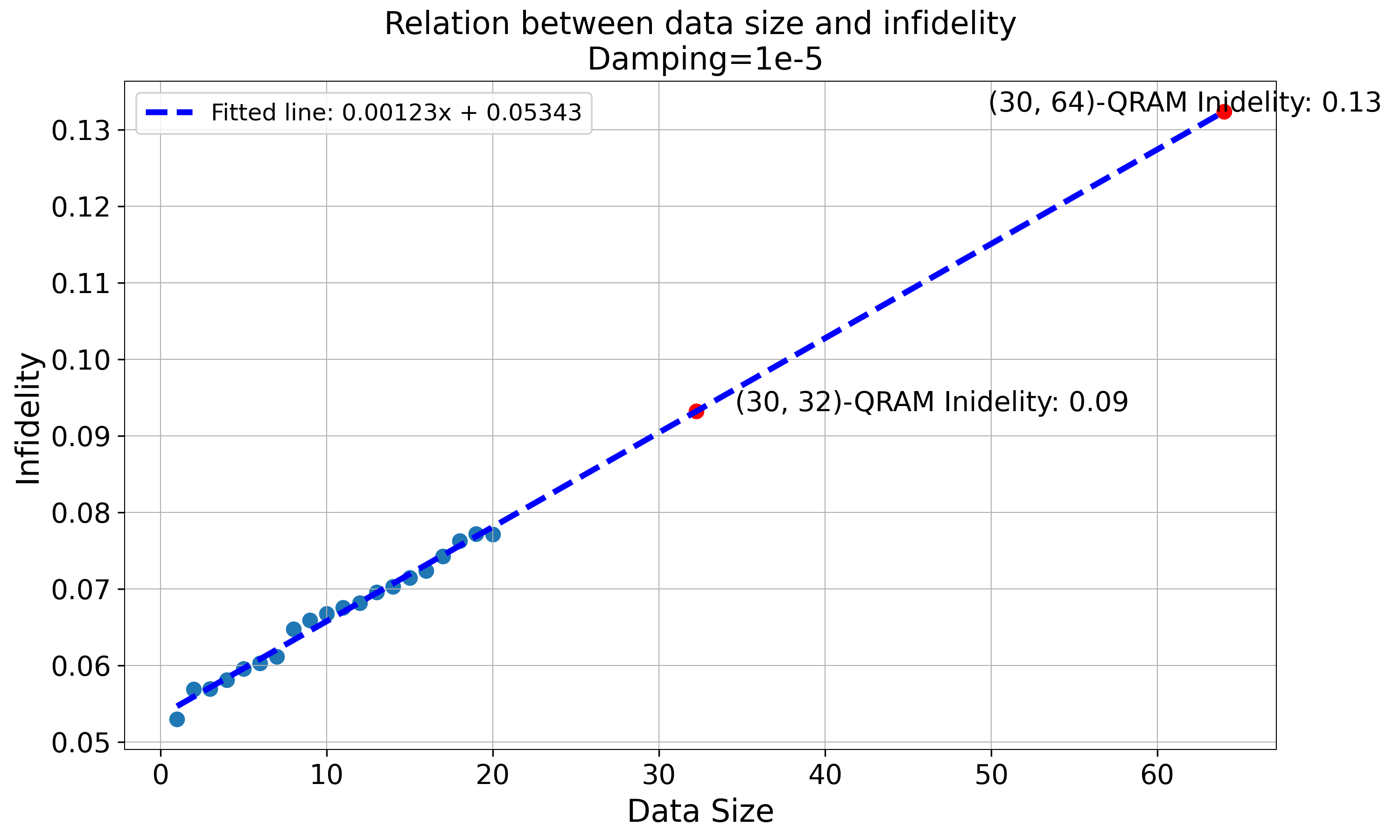}
    \caption{The figure presents the fitting function and annotates the predicted fidelities of $(30, 32)$-QRAM and $(30, 64)$-QRAM.}
    \label{fig:Fitting_data_full_64_damping}
\end{figure}

We leveraged the data obtained from these simulations to derive polynomial expressions that accurately capture the relationship between infidelity and address size. Subsequently, these quadratic expressions were utilized to extrapolate and predict the infidelities of $(30, k)$-QRAM for a range of $k$ values from $1$ to $20$. As a result, we have obtained a comprehensive set of predicted QRAM infidelities for $(30, k)$ configurations, where $k$ ranges from $1$ to $64$, as shown in  Figure~\ref{fig:Fitting_data_full_64_damping}.

\end{document}